\documentclass[twoside,twocolumn,9pt]{article}
\usepackage[version=3]{mhchem}
\usepackage[left=1.5cm, right=1.5cm, top=1.785cm, bottom=2.0cm]{geometry}
\usepackage{balance}
\usepackage{mathptmx}
\usepackage{mathtools}
\usepackage{sectsty}
\usepackage{graphicx} 
\usepackage{lastpage}
\usepackage[format=plain,justification=justified,singlelinecheck=false,font={stretch=1.125,small,sf},labelfont=bf,labelsep=space]{caption}
\usepackage{float}
\usepackage{fancyhdr}
\usepackage{fnpos}
\usepackage[english]{babel}
\addto{\captionsenglish}{%
  
}
\usepackage{array}
\usepackage{droidsans}
\usepackage{charter}
\usepackage[T1]{fontenc}
\usepackage[usenames,dvipsnames]{xcolor}
\usepackage{setspace}
\usepackage[compact]{titlesec}
\usepackage{hyperref}
\usepackage{epsfig}
\usepackage{subfig}
\usepackage{adjustbox}
\usepackage{color}
\usepackage{authblk}
\usepackage{epstopdf}%This line makes .eps figures into .pdf - please comment out if not required.
\title{Machine learning opens a doorway for microrheology with optical tweezers in living systems}

\author[a]{Matthew G. Smith}
\author[b]{Jack Radford} 
\author[c]{Eky Febrianto}
\author[d]{Jorge Ramírez}
\author[a]{Helen O'Mahony}
\author[e]{Andrew B. Matheson}
\author[b]{Graham M. Gibson} 
\author[b]{Daniele Faccio} 
\author[a,1]{Manlio Tassieri}

\affil[a]{Division of Biomedical Engineering, James Watt School of Engineering, University of Glasgow, Glasgow G12 8LT, U.K.}
\affil[b]{School of Physics and Astronomy, University of Glasgow, Glasgow G12 8QQ, U.K.}
\affil[c]{Glasgow Computational Engineering Centre, James Watt School of Engineering, University of Glasgow, Glasgow G12 8LT, U.K.}
\affil[d]{Departamento de Ingenier\'{i}a Qu\'{i}mica Industrial y Medio Ambiente, Universidad Polit\'{e}cnica de Madrid, Jos\'{e} Guti\'{e}rrez Abascal 2,28006 Madrid, Spain}
\affil[e]{School of Engineering and Physical Sciences, Institute of Biological Chemistry, Biophysics and Bioengineering, Heriot Watt University, Edinburgh, U.K.}

\begin{document}

\maketitle
\begin{abstract}
It has been argued [Tassieri, \textit{Soft Matter}, 2015, \textbf{11}, 5792] that linear microrheology with optical tweezers (MOT) of living systems ``\textit{is not an option}'', because of the wide gap between the observation time required to collect statistically valid data and the mutational times of the organisms under study.
Here, we have taken a first step towards a possible solution of this problem by exploiting modern machine learning (ML) methods to reduce the duration of MOT measurements from several tens of minutes down to one second.
This has been achieved by focusing on the analysis of computer simulated trajectories of an optically trapped particle suspended in a set of Newtonian fluids having viscosity values spanning three orders of magnitude, i.e. from $10^{-3}$ to $1$ Pa$\cdot$s.
When the particle trajectory is analysed by means of conventional statistical mechanics principles, we explicate for the first time in literature the relationship between the required duration of MOT experiments ($T_m$) and the fluids relative viscosity ($\eta_r$) to achieve an uncertainty as low as $1\%$; i.e., $T_m\cong 17\eta_r^3$ minutes.
This has led to further evidences explaining why conventional MOT measurements commonly underestimate the materials' viscoelastic properties, especially in the case of high viscous fluids or soft-solids such as gels and cells. Finally, we have developed a ML algorithm to determine the viscosity of Newtonian fluids that uses feature extraction on raw trajectories acquired at a kHz and for a duration of only one second, yet capable of returning viscosity values carrying an error as low as $\sim0.3\%$ at best; hence the opening of a doorway for MOT in living systems.
\end{abstract}

\section*{Scientific Significance}
Optical tweezers have been successfully adopted as exceptionally sensitive transducers for microrheology studies of time-invariant complex fluids. However, in contrast to a general practice, in this article we provide experimental evidences elucidating that a similar approach should not be adopted for living systems because the observation time required to acquire statistically valid measurements is much longer than the shortest mutational time of the organism under study, with a consequent violation of the fluctuation–dissipation theorem underpinning passive microrheology methods. Here we provide a first step towards the solution of this problem by adopting modern machine learning methods to shorten the measurement time down to 1 second, thus offering the opportunity to perform microrheology with optical tweezers in living systems.

\section*{Introduction}
Since their first appearance in the 1970s \cite{Ashkin1970,Ashkin1971,Ashkin1986}, Optical Tweezers (OT) have been employed as extremely sensitive force transducers across a variety of disciplines within the \textit{Natural Sciences} \cite{Svoboda1994,Diekmann2016,Davies2016,Ayala2016,Weigand2017,Kim2016}.
OT rigs rely on the ability of a highly focused laser beam to optically trap in 3D micron sized dielectric particles suspended in a fluid. This is achieved by optically guiding a monochromatic laser beam through a microscope objective with a high numerical aperture.
Once trapped, the particle experiences a quadratic potential and therefore a restoring force that is linearly proportional to the distance of the particle from the trap centre; with a constant of proportionality of the order of a few $\mu$N/m.
It follows that by measuring the particle position to a high spatial resolution (i.e., of a few nm), scientists have successfully measured forces as low as a few pNs, such as those generated by the thermally driven motion of water molecules \cite{Madsen2021} or those exerted by single motor proteins \cite{Capitanio2012}.
Interestingly, accessing particles' trajectory to high temporal and spatial resolutions is one of the requirements underpinning microrheology techniques \cite{Squires2010,Waigh2005}, as elucidated in this paper for the specific case of Optical Tweezers.

Microrheology is a branch of rheology (the study of the flow of matter) and is focused on the characterization of the mechanical properties of complex materials by performing measurements at micron length scales, often with sample volumes as little as a few microlitres.
This offers an indisputable advantage over classical bulk rheology techniques, which require millilitres of sample, especially in biophysical studies where samples are often rare and/or precious and come in small quantities (e.g., a few micro-litres).
Microrheology techniques are categorised as either `passive' or `active' depending on whether the motion of the tracer particles is thermally driven or induced by an external force field, respectively.
Interestingly, optical tweezers is one of such techniques that can be defined as a \textit{hybrid} microrheology tool \cite{Tassieri2016Book}, because of the quadratic nature of the optical potential constraining the motion of the probe particle. Indeed, despite the tracer particles being optically trapped (within the focal plane of a microscope), at short time scales (i.e., for small displacements) the restoring force exerted on the probe is weak enough for the particle to experience Brownian motion because of the thermal fluctuations of the molecules of the suspending media. Nonetheless, active microrheology with OT is still possible by driving the trapping laser, often in a sinusoidal pattern as elucidated within Refs \cite{Valentine1996,Tassieri2015_MOT}.
However, as we shall further corroborate in this work, a necessary condition for executing either passive or active linear MOT measurements is to perform ``\textit{sufficiently}'' long measurements, commonly of the order of tens of minutes \cite{Tassieri2015_MOT,Tassieri2019}.
This is because most of the analytical methods used to determine the materials' viscoelastic properties are underpinned by statistical mechanics principles, whose accuracy relies on the analysis of a \textit{significant} number of independent readings.
Therefore, as pointed out by Tassieri \cite{Tassieri2015_MOT}, it may not be appropriate to adopt MOT for studies involving living systems, as biological processes occur at time-scales ranging from $10^{-2}$ to $10^{2}$sec \cite{Harrison2013,Mizuno2007,Toyota2011,Alberts2015}, and therefore the viscoelastic properties of biological systems may not be considered time-invariant during the measurements.

Hence, the aim of this work to exploit modern Machine Learning (ML) methods to reduce the duration of MOT measurements and thus allow scientists to perform microrheology measurements in living systems.
In order to achieve such a challenging aim, in this work, we have taken a first step towards a possible solution of the problem by focusing on the analysis of computer simulated trajectories of an optically trapped particle suspended within a set of Newtonian fluids having viscosity values spanning three decades, i.e. from $10^{-3}$ to $1$ Pa$\cdot$s.
The goal was to develop a ML algorithm that would effectively estimate fluids' viscosity from relatively short measurements ($\leq 1$sec) and compare the outcomes with those obtained by analysing the same set of data with conventional methods based on statistical mechanics principles \cite{Tassieri2010,Tassieri2012,Tassieri2015_Glance,Tassieri2019}.
Notably, this study has led to the following key findings: (i) we corroborate the requirement for MOT studies to perform ``\textit{sufficiently}'' long measurements when using conventional analytical methods for data analysis and (ii) we provide, for the first time in literature, a means for estimating the required duration of the experiment to achieve an uncertanty as low as $1\%$; (iii) we provide evidence explaining why conventional MOT measurements commonly underestimate the materials' viscoelastic properties, especially in the case of high viscous fluids or soft-solids (e.g., gels and cells); (iv) we have developed a ML algorithm that uses feature extraction on only `one second' of trajectory data to determine the viscosity of Newtonian fluids; yet capable of returning viscosity values carrying an error as low as $\sim0.3\%$ at best and of $\sim7\%$ at worst, which is five times smaller than those obtained from conventional analytical methods applied to the same data.

\begin{figure*}[t!]
    \centering
    {\includegraphics[trim={0 0 0 0},clip,width=18 cm]{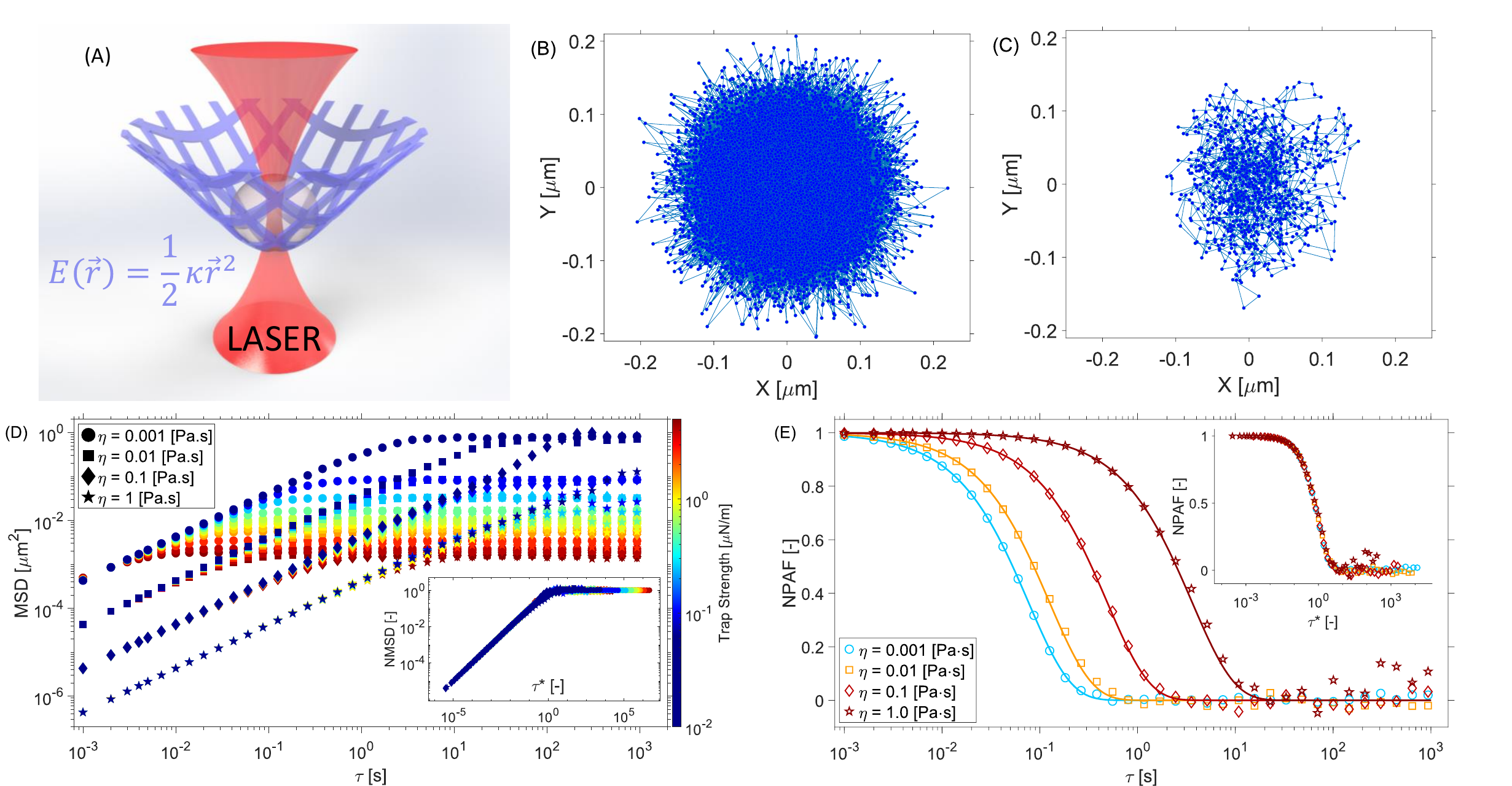}}
\caption{(A) A schematic representation of an optically trapped bead within a harmonic potential, $E(\vec{r})$, where $\kappa$ is the trap stiffness and $\vec{r}$ is the bead position from the trap centre. (B-C) Two examples of 2D trajectories of an optically trapped bead of radius of $1\mu$m suspended in water for $1024$s (B) and for $1$s (C). Both trajectories were generated by means of a MATLAB code adapted from the one developed by Volpe \cite{Volpe2013}. (D) The mean square displacement (MSD) curves of a series of $48$ simulated trajectories of $1024$ sec duration and acquired at a $1$ kHz of an optically trapped particle experiencing constraining forces ranging from $0.01$ to $5$ $\mu$N/m (see colour bar) and suspended into four different Newtonian fluids having viscosity values spanning three orders of magnitude (see legend). The inset shows the same data as in the main, but with the ordinate axis normalised by the twice the variance of the particle trajectory and the abscissa $\tau$ replaced by the dimensionless lag-time $\tau^*$, as elucidated in the body of the manuscript. (E) Four examples of normalised position autocorrelation functions (NPAF, symbols) of a particle suspended in four Newtonian fluids having viscosity of $10^{-3}$, $10^{-2}$, $0.1$, $1$Pa$\cdot$s and experiencing constraining forces of $0.25$, $1.5$, $4$, $5\mu$N/m (from left to right and colour coded as for the colour bar in D), respectively. The lines are single exponential decay functions $A(\tau)=e^{-\lambda\tau}$ drawn with $\lambda=\kappa/(6\pi\eta a)$ evaluated by using the input (nominal) parameters mentioned above; i.e., $\lambda=13.26$, $7.96$, $2.12$ and $0.27$Hz, respectively. The inset shows the same data as in the main (symbols only), but the abscissa has been replaced by $\tau^*=\lambda\tau$.}
\label{fig:Positional Data}
\end{figure*}

\section*{Theoretical Background}

\subsection*{Passive Microrheology with Optical Tweezers}
Passive MOT is typically performed by means of a stationary optical trap that confines in 3D a spherical particle suspended in a fluid of unknown viscoelastic properties. At thermal equilibrium, the Brownian motion of the probe particle is caused by the thermal fluctuations of the fluids' molecules and it is monitored by means of a high speed motion detection device. The particle trajectory is typically extracted in 2D, as the one shown in Fig. \ref{fig:Positional Data}(B-C).
Notably, a statistical mechanics analysis of the particle's trajectory can return not only the trap stiffness of the OT, but also a good estimation of the frequency-dependent viscoelastic properties of the suspending fluid \cite{Tassieri2010,Preece2011,Tassieri2012,Pommella2013,Lee2014,Tassieri2015_Glance,Tassieri2019,Smith2021}. The latter can be evaluated by solving a generalised Langevin equation as the following one:
\begin{equation}
    m\vec{a}(t)=\vec{f_R}(t)-\int_{0}^{t}\xi(t-\tau)\vec{v}(\tau)d\tau-\kappa\vec{r}(t),
    \label{eqn:Langevin}
\end{equation}
where $m$ is the mass of the particle, $\vec{a}(t)$ is its acceleration, $\vec{v}(t)$ is its velocity, $\vec{r}(t)$ is its position, $\vec{f_R}(t)$ is the Gaussian white noise term used for modelling the stochastic thermal forces, and $\xi(t)$ is the generalised time-dependent memory function accounting for the viscoelastic nature of the fluid \cite{Mason1995}. The convolution integral represents the time-dependent friction force exerted by the complex fluid onto the particle. The term $\kappa\vec{r}(t)$ is the restoring force of the optical trap, when the confining field $E(\vec{r})$ exerted by the optical tweezers is assumed to have an harmonic form:
\begin{equation}
\label{eqn:Potential}
    E(\vec{r}) = \frac{1}{2}\kappa \vec{r}^2,
\end{equation}
where $\kappa$ is the trap stiffness and $\vec{r}$ is the particle position from the trap centre.
Interestingly, in the case of Newtonian fluids (i.e., for purely viscous fluids with constant viscosity $\eta$) and at low Reynolds numbers (for which the inertia term can be neglected), Equation \ref{eqn:Langevin} simplifies as follows \cite{Volpe2013}:
\begin{equation}
   \vec{W}(t)\sqrt{2k_BT\gamma}=\gamma\vec{v}(t)+\kappa\vec{r}(t),
    \label{eqn:SimplifiedLangevin} 
\end{equation}
where the term on the left side represents the fluctuating force due to random impulses from many neighboring fluid molecules, $\gamma=6\pi\eta a$ is the friction coefficient, $a$ is the particle radius, $k_B$ is the Boltzmann's constant, and $T$ is the absolute temperature.
In this work, Equation \ref{eqn:SimplifiedLangevin} has been adopted to generate (thousands of) 2D trajectories of optically trapped particles suspended into a set of Newtonian fluids having different viscosity values for machine learning purposes, as explained in the following sections.

\begin{figure*}[t!]
    \centering
    {\includegraphics[trim={0 0 0 0},clip,width=\textwidth]{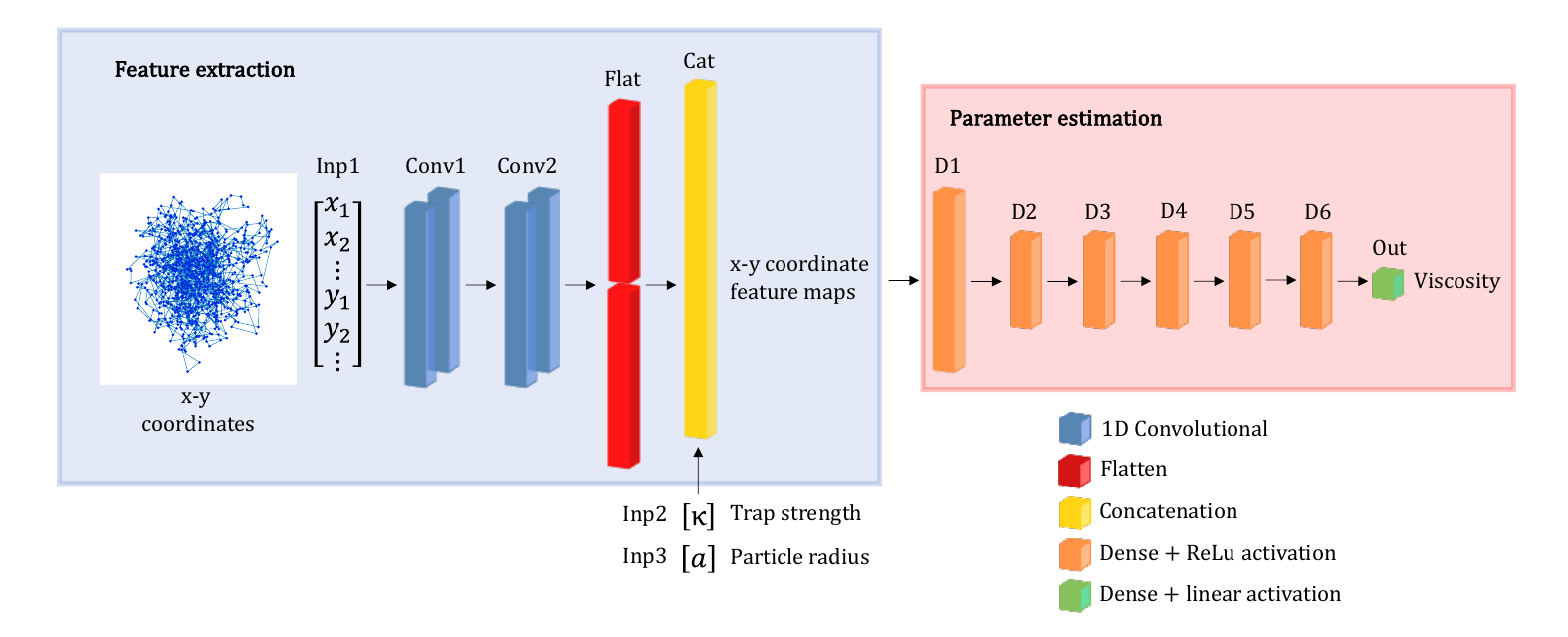}}
    \caption{A schematic representation of the machine learning architecture used in this work. A single particle trajectory of x-y coordinates is transformed and preprocessed for feature extraction. The output of which is concatenated with the trap strength ($\kappa$) and the particle radius ($a$). This is used as the input for parameter estimation with a single output node of viscosity.}
    \label{fgr:ML_Arch}
\end{figure*}

In the general case, i.e. for any generic complex fluid, it has been shown \cite{Tassieri2010,Preece2011,Tassieri2012} that Equation \ref{eqn:Langevin} can be solved for the fluids' complex shear modulus ($G^*(\omega)$) in terms of either of the particle normalised mean squared displacement (NMSD), $\Pi(\tau)$, or its normalised position autocorrelation function (NPAF), $A(\tau)$; which are both drawn in the insets of Figure \ref{fig:Positional Data}(D, E) for some of the cases studied in this work. Notably, these two functions are simply related to each other and their expressions are:
\begin{equation}
    \Pi(\tau)=\frac{\langle \Delta r^2(\tau)\rangle_{t_0}}{2\langle r^2\rangle_{eq.}}\equiv \frac{\langle [r(t_0+\tau)-r(t_0)]^2\rangle_{t_0}}{2\langle r^2\rangle_{eq.}} =1-A(\tau),
    \label{eqn:NMSD}
\end{equation}
where $\tau$ is the lag-time ($t-t_0$) and the brackets $\langle...\rangle_{t_0}$ represent an average over all initial times $t_0$.
The relationship between the above two time-averaged functions and the time-invariant fluids' complex shear modulus is:
\begin{equation}
\label{G*OT}
	G^*(\omega)\frac{6\pi a}{\kappa}=\left(\frac{1}{i\omega \hat{\Pi}(\omega)}-1\right)\equiv \left(\frac{1}{i\omega\hat{A}(\omega)}-1\right)^{-1}\equiv \frac{\hat{A}(\omega)}{\hat{\Pi}(\omega)},
\end{equation}
where $\hat{\Pi}(\omega)$ and $\hat{A}(\omega)$ are the Fourier transforms of $\Pi(\tau)$ and $A(\tau)$, respectively. The inertial term ($m\omega^{2}$) present in the original works\cite{Tassieri2010,Preece2011} has been here neglected, because for micron-sized particles it only becomes significant at frequencies of the order of MHz.

Notably, in the case of Newtonian fluids, the above equations simplify significantly and the relationship between the fluids' viscosity and the particle trajectory reads as follows:
\begin{equation}
\label{eqn:etaNew}
    \Pi(\tau)=1-A(\tau)=1-e^{-\lambda\tau},
\end{equation}
where $\lambda=\kappa/(6\pi\eta a)$ is the characteristic relaxation rate (also known as the ``\textit{corner frequency}''\cite{Berg-Sorensen2004}) of the compound system OT \textit{plus} fluid.
Notably, it has been shown \cite{Tassieri2015_Glance} that by plotting $\Pi(\tau)$ and $A(\tau)$ versus a dimensionless lag-time $\tau^*=\tau\lambda$, all the curves would collapse onto a master curve, as shown in the insets of Figure \ref{fig:Positional Data} (D) and (E), respectively.
It follows that, for Newtonian fluids it is a straightforward step to determine their viscosity by analysing the temporal behavior of the NPAF \cite{Tassieri2015_Glance}.
In particular, by plotting the natural logarithm of $A(\tau)$ versus $\tau$, one would obtain a straight line having a slope equal to $-\lambda$, from which the viscosity could be determined by means of a simple linear fit.
In this work the fitting procedure has been constrained to the ordinate values ranging from $0$ and $-1$ (equivalent to $A(\tau)=1$ and $A(\tau)=e^{-1}$, respectively) to minimize the error, as discussed hereafter.

\begin{figure*}[t!]
    \centering
    {\includegraphics[trim={0 0 0 0},clip,width=\textwidth]{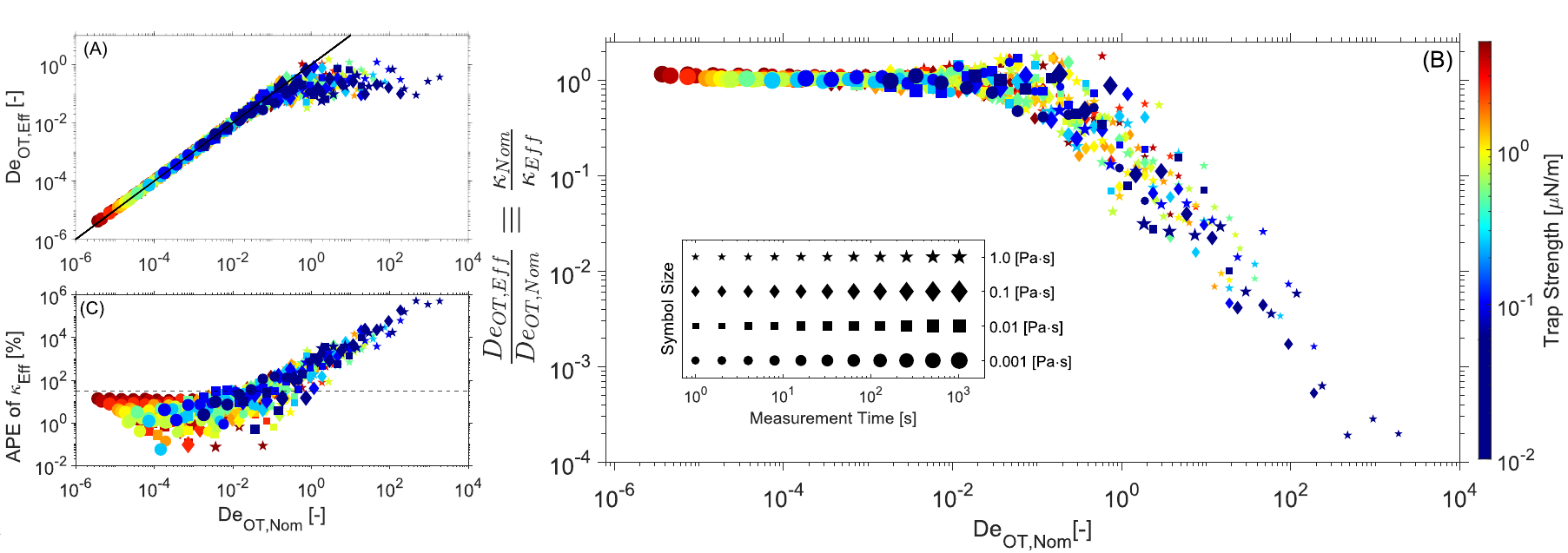}}
    \caption{(A) The effective Deborah number for optical tweezers $De_{OT,Eff.}$ \textit{vs}. the nominal one $De_{OT,Nom.}$ for a series of $528$ simulated trajectories of different duration of an optically trapped particle experiencing various constraining forces and suspended into four different Newtonian fluids having viscosity values spanning three orders of magnitude. (B) The same data as in (A), but rearranged as the ratio between the two Deborah numbers \textit{vs}. $De_{OT,Nom.}$ to highlight the transition from calibrated measurements to miscalibrated ones occurring at $De_{OT,Nom.}\sim 1$. (C) The absolute percentage error (APE) of $\kappa_{Eff.}$ \textit{vs}. $De_{OT,Nom.}$ for the same set of trajectories as in (A). The line indicates an APE value of $30\%$, as reported in Ref.\cite{Matheson2021}. The colour bar indicates trap stiffness used during the generation of the trajectories. The size of the symbols scales with the measurement time as shown in the inset of (B).}
    \label{fgr:Calibration}
\end{figure*}

\section*{Methods}
\subsection*{Simulation of particle trajectories}
In order to train and test the machine learning algorithm discussed in the next section, we have used Equation \ref{eqn:SimplifiedLangevin} to generate thousands of trajectories by means of a MATLAB code adapted from the one developed by Volpe \cite{Volpe2013}, which is able to simulate a 2D trajectory of an optically trapped particle suspended into a Newtonian fluid. The input parameters of the code were the trap stiffness, the viscosity, the temperature, the particle radius, the acquisition rate and the number of individual readings required. For instance, in Figure \ref{fig:Positional Data} are shown two examples of trajectory having the same input parameters, but duration of $10^3$s in (B) and $1$s in (C).

Moreover, in order to investigate the impact of the measurements duration on the outcomes obtained from both the conventional and the ML enhanced MOT approaches, we generated a set of particle trajectories suspended into four different Newtonian fluids having viscosity values of $10^{-3}$, $10^{-2}$, $0.1$ and $1$ Pa$\cdot$s, respectively; and trap strengths ranging from $0.01$ to $5$ $\mu$N/m. These trajectories were simulated for $1024$ sec at an acquisition rate of $1$ kHz, which is equivalent to a real measurement of circa $17$ minutes in duration.
Notably, due to their stochastic nature, it is possible to split each of these trajectories into shorter ones of variable duration, down to $0.05$ sec.
All these trajectories were analysed to calculate the fluids' viscosity by means of Equation \ref{eqn:etaNew}, and the mean absolute percentage error (MAPE) of the outcome was calculated for each trajectory by means of the following equation:
\begin{equation}
    MAPE = \frac{100}{N}\sum_{i=1}^{N}\Bigg| \frac{\eta_i-\eta_{0i}}{\eta_{0i}}\Bigg|,
\end{equation}
where $N$ is the number of trajectories for a given duration, $\eta_{0}$ is the nominal viscosity value (used as input in the simulations) and $\eta$ is the measured one.

\subsection*{Machine Learning architecture}
In fluid mechanics, machine learning (ML) has been widely used to translate observational and experimental data into knowledge about the underlying physics of the fluid~\cite{Brunton2020}. Depending on the information being used for learning, ML algorithms can be categorised into supervised, semisupervised, and unsupervised. In this work we consider a supervised ML algorithm where the input (i.e., the particle trajectories) and the respective output (i.e., the viscosity) are used during learning. Specifically, we consider feed-forward neural networks (NNs), or multilayer perceptrons~\cite{Hornik1989, Bishop2006}, as the nonlinear function approximations between the input and output. The standard feed-forward NNs passes the input information through a network of hidden units and activation functions to produce the prediction. Deep Neural Networks (Deep NN)~\cite{LeCun2015, Goodfellow2016}  obtains a nonlinear approximation through the composition of multiple hidden layers.
To obtain the unknown network weights, nonlinear optimisation methods, such as backpropagation~\cite{Rumelhart1986}, are used by minimising the discrepancy between the predictions and the known training outputs.

In this paper, we sidestep the conventional method (i.e., Equation \ref{eqn:etaNew}) of estimating fluids' viscosity from the trajectories of optically trapped particles by means of supervised ML. The training dataset consists of $100,000$ particle trajectories, each of $10$s duration, for different fluids' viscosity. 

In order to cover the range of explored viscosity (i.e., from $0.001$ to $1$ Pa$\cdot$s), the viscosity values are randomly sampled from a log-uniform distribution ranging from $0.0008$ to $1.2$ Pa$\cdot$s. Similarly, the trap strengths are randomly sampled from a uniform distribution ranging from $0.08$ to $0.39$ $\mu$N/m. Prior to the training, each trajectory coordinate input is flattened into a one-dimensional array to better consider the trajectory's temporal correlation. Moreover, we also consider in this work shorter observation times of the trajectories, i.e., $T_m = \{1\text{s}, \, 0.5\text{s}, \, 0.1\text{s}, \, 0.05\text{s}\}$, which are obtained through subdivision of the original $10$s datasets. 

\begin{figure*}[t!]
    \centering
    \includegraphics[trim={0 0 0 0},clip,width=1\textwidth]{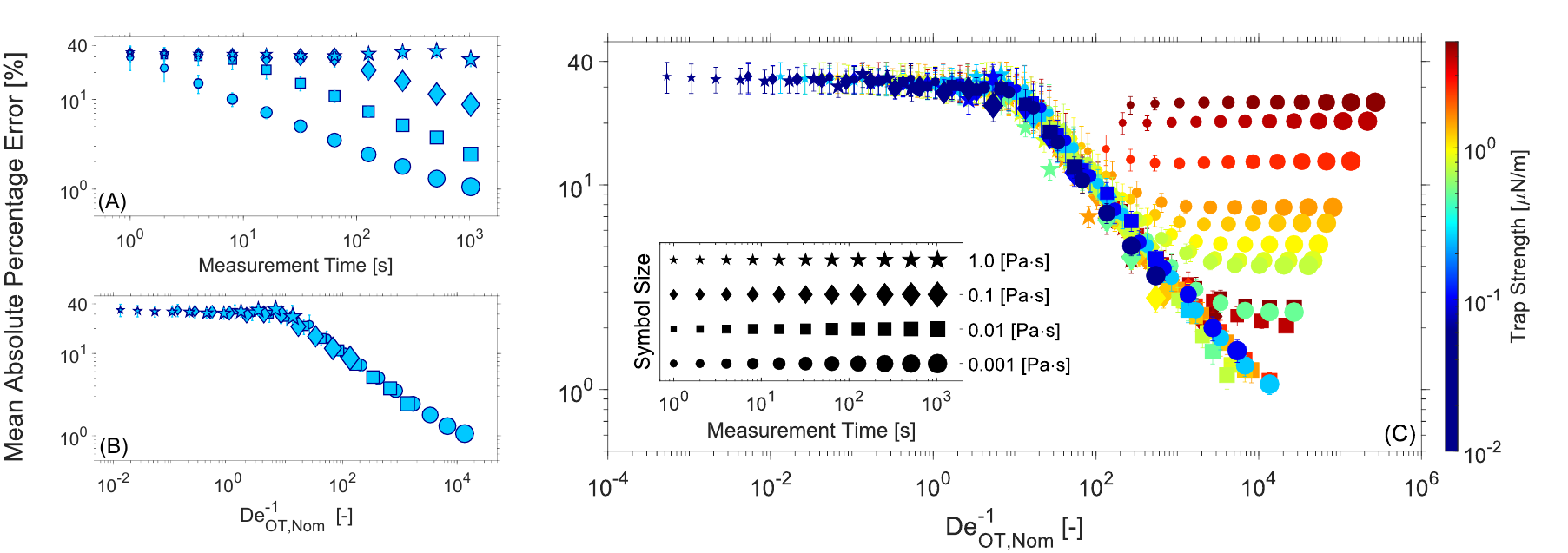}
    \caption{(A) The mean absolute percentage error (MAPE) of viscosity \textit{vs}. measurement duration ($T_m$) determined by using the analytical method described in the body of the manuscript. (B) The same data as in (A), but drawn \textit{vs}. the inverse of the nominal Deborah number for optical tweezers ($De_{OT,Nom.}$), which is proportional to the measurement duration. (C) The MAPE \textit{vs}. $De^{-1}_{OT,Nom.}\propto T_m$ for a series of $528$ simulated trajectories of different duration of an optically trapped particle experiencing various constraining forces and suspended into four different Newtonian fluids having viscosity values spanning three orders of magnitude. All three diagrams have the same ordinate. The colour bar indicates the nominal trap stiffness used during the generation of the trajectories. The size of the symbols scales with $T_m$ as shown in the inset of (C).}
    \label{fig:MAPE}
\end{figure*}

%Architecture - CNN
Figure 2 shows a schematic representation of the ML architecture used in this study, consisting of two blocks, i.e., feature extraction and parameter estimation. The one-dimensional input, obtained by flattening the coordinates, is first processed through the feature extraction block, which comprises one-dimensional convolutional neural network (CNN) layers~\cite{LeCun1989, LeCun1995}. The CNN acts as a convolution operator that enhances local temporal structures appearing in the particle trajectories.
The resulting `feature maps' are transformations of the input data into latent variables which highlight important information for the task of estimating viscosity. We consider in this study two CNN layers with increasing filter widths of $10$ and $100$, respectively. These are equivalent to $0.01$s and $0.1$s windows of the particle trajectory.
Note that, for observation time $T_m=0.1$s and $T_m=0.05$s, the second convolutional layer filter sizes were adjusted, due to the shorter vector lengths, to be $50$ and $25$, respectively. 

%Architecture - NN
The feature maps are then concatenated, along with the trap stiffness and particle radius, to a 1D vector and passed to six fully-connected hidden layers and an output layer which predicts a single value of viscosity.
The concatenation of additional variables is crucial to discriminate between fluids which have different viscosity, but similar particle trajectories due to other dependant variables (e.g., trap stiffness and particle radius). Each neuron in the hidden layers uses a ReLu (rectified linear unit) activation function, while a linear activation function is used in the output layer. The loss function was chosen to be the mean absolute percentage error (MAPE) to prevent bias in training towards minimising losses for high viscosity values with larger residuals. The hyper-parameters of the model including the batch size, learning rate, number of epochs and validation split were $256$, $10^{-5}$, $200$ and $0.1$, respectively.
The training has been performed in triplicate for each model with input trajectories having interval $T_m$ from $0.05$ to $1$ seconds, using an Adam optimizer~\cite{Kingma2014}, and was performed on a desktop PC equipped with an $18$-core Intel i9-10980XE CPU ($3$GHz), $256$GB RAM and an NVIDIA GeForce RTX $3090$ with $24$GB memory. The training time for each ML model increased with decreasing input length due to the increasing number of training examples so each model took from $2.5-6.5$hrs to train depending on input trajectory.

\section*{Results and Discussion}
One of the key features and advantages of using optical tweezers for microrheology purposes is that they can be easily calibrated without the use of external transducers.
Indeed, as we shall discuss hereafter, it has been assumed \cite{Tassieri2010,Preece2011,Tassieri2012,Pommella2013,Lee2014,Tassieri2015_Glance,Tassieri2019} that the trap stiffness of symmetric OT can be determined to a high accuracy by appealing to the principle of equipartition of energy:
\begin{equation}
\label{eqn:EquiEn}
    \frac{d}{2}k_BT= \frac{1}{2}\kappa \langle \vec{r}^2\rangle_{eq.},
\end{equation}
where $d$ is the dimension of the motion.
As we shall demonstrate hereafter, this is true as long as the measurement time is ``\textit{sufficiently}'' longer than the characteristic time $\tau_{OT}$ of the compound system made of OT (i.e., its trap stiffness), fluid (i.e., its compliance) and bead (i.e., its radius taken as a characteristic length of the probe), which is not known \textit{a priori} in rheological investigations of complex materials.

However, in the case of Newtonian fluids and operational condition of the instrument within the micro length- and time-scales, as mentioned earlier, the compound system has a single characteristic time defined as $\tau_{OT}=\lambda^{-1}$, which can be used as a reference to estimate the minimum measurement duration required to properly calibrate the trap stiffness.
In particular, by defining the duration of a measurement ($T_m$) as the ratio between the total number of readings ($N$) and the acquisition rate ($f=samples/s$) of the detector used for tracking the particle position, one could define a Deborah number \cite{Reiner1964} for optical tweezers ($De_{OT}$) as:
\begin{equation}
\label{eqn:De}
De_{OT}=\frac{\tau_{OT}}{T_m}=\frac{6\pi\eta af}{N\kappa},
\end{equation}
which can be further differentiated into ``\textit{nominal}'' ($De_{OT,Nom.}$) and ``\textit{effective}'' ($De_{OT,Eff.}$), depending on whether the trap stiffness used for determining $\lambda$ is the nominal value set as input in the simulation code generating the trajectories or the measured one by means of Equation \ref{eqn:EquiEn}, which is affected by $T_m$ as demonstrated hereafter.

In Figure \ref{fgr:Calibration}--(A) we report $De_{OT,Eff.}$ \textit{vs}. $De_{OT,Nom.}$ for a series of $528$ simulated trajectories of variable duration of an optically trapped particle experiencing various constraining forces and suspended in four different Newtonian fluids having viscosity values spanning three orders of magnitude.
From Figure \ref{fgr:Calibration}--(A) it is apparent the existence of a crossover value of $De_{OT,Nom.}\sim 1$ delimiting two operating ranges of OT rigs, i.e.: (i) for $De_{OT,Nom.}<<1$, the trap stiffness is determined to a high accuracy via Equation \ref{eqn:EquiEn}; whereas, (ii) for $De_{OT,Nom.}>>1$, the constraining force is \textit{undetermined}.
In particular, as it often happens in many real experiments for which $T_m$ is not sufficiently long, $\kappa_{Eff.}$ is \textit{overestimated} as shown in Figure \ref{fgr:Calibration}--(B), where the same data as in Figure \ref{fgr:Calibration}--(A) are reported as the ratio between the two Deborah numbers (equivalent to the ratio between the two values of the trap stiffness $\kappa_{Nom.}/\kappa_{Eff.}$) vs. $De_{OT,Nom.}$.
Based on Equation \ref{G*OT}, it follows that when the trap stiffness is \textit{overestimated} the outcomes of MOT measurements are \textit{underestimated}, especially when they are attempted in high viscous fluids or soft-solids (e.g., gels and cells) \cite{Ashworth2020}.
Our findings are further corroborated by the data shown in Figure \ref{fgr:Calibration}--(C), where the absolute percentage error (APE) of $\kappa_{Eff.}$ is reported against $De_{OT,Nom.}$.
Notably, the horizontal line reported in Figure \ref{fgr:Calibration}--(C) represents the threshold value of the APE of $\kappa_{Eff.}$ identified experimentally by Matheson \textit{et al}. \cite{Matheson2021} below which microrheology measurements performed with OT return an APE of the fluids' viscosity lower than circa $5\%$ (see Figure 5 of Ref.\cite{Matheson2021}). Moreover, the experimental evidence reported by Matheson \textit{et al}. \cite{Matheson2021} is in agreement with the results reported in this work, as elucidated in the following paragraphs.
It follows that, based on a conservative approach, one could argue that only for $De_{OT,Eff.}\leq0.001$ an accurate calibration could be achieved; which implies a minimum measurement duration of $T_m\geq1000\times\tau_{OT}$ for a given system.

For instance, in the case of two measurements both performed at room temperature (i.e., $T=20^o$C) with a bead of $1\mu$m in radius and a trap stiffness of $\kappa=2\mu$N/m, but one in water (with $\eta=0.001$Pa$\cdot$s) and the other in a fluid having a viscosity thousand times higher than water (e.g., glycerol), the characteristic times of the two compound systems would be $\tau_{OT}\cong 0.01$s and $\tau_{OT}\cong 10$s, respectively.
It follows that, in order to achieve an accurate calibration of the OT (i.e. for $De_{OT}\leq0.001$), the measurements should last at least $10$s and $2.78$hrs, respectively.

At this point it is important to highlight that optical tweezers rigs are commonly equipped with either a camera or a quadrant photodiode (QPD) device for tracking the particle position to a high acquisition rate, often operating at KHz or MHz, respectively.
It follows that, when microrheology measurements are performed on materials with a higher viscous character than water, significantly longer measurements would be required, and therefore rigs equipped with either a QPD or an ultra high-speed camera would be more prone to be miscalibrated. This is because they are often equipped with an insufficient capacity of random access memory (RAM) to process the high--volume of data (of several MB/s) generated during the particle tracking procedure (of possible duration of $T_m=10^4$s, which would result in >10 GB RAM occupancy); thus, they either crash or, in order to avoid this, measurements are stopped early causing a $De_{OT}>>1$.
A possible solution to avoid memory clogging, but not the lengthiness of measurements, is achieved by equipping the rig with an online digital correlator, which allows the machine to process high-volume data streams and to compress the relevant information in real-time, thus minimising the use of RAM \cite{Yanagishima2011, Ramirez_2010}.

\begin{figure}[t]
    \centering
    \includegraphics[width=9 cm]{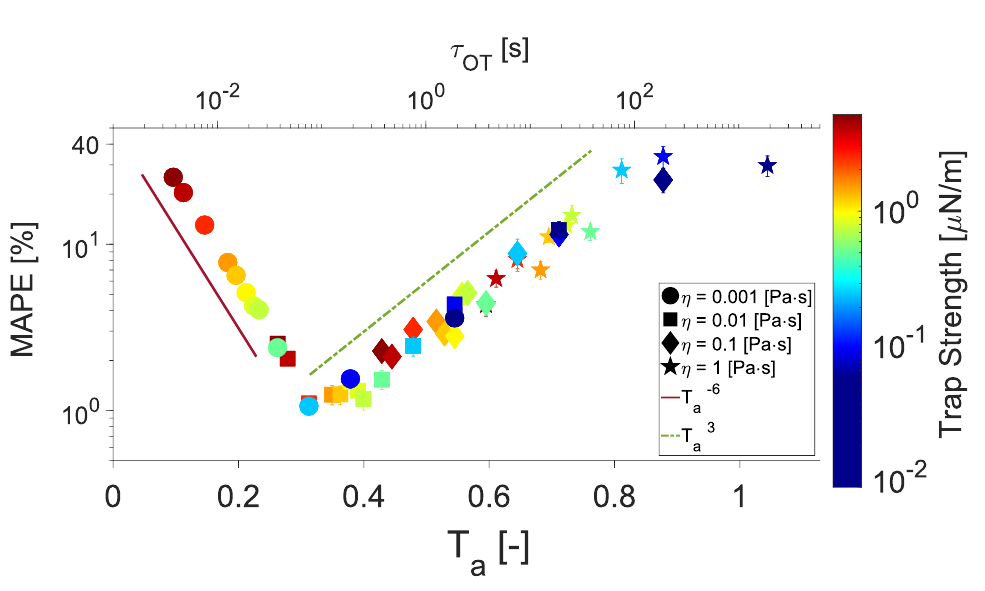}
    \caption{Data taken from Figure \ref{fig:MAPE} for the trajectories with maximum $T_m$ and drawn \textit{vs}. $T_a$. The top axis reports the characteristic time of the compound system $\tau_{OT}$. The colour bar indicates the nominal trap stiffness used during the generation of the trajectories. The two lines are guides for the power laws as indicated in the legend.}
    \label{fig:TaNumber}
\end{figure}

Let us now investigate how $T_m$ affects the evaluation of the fluid viscosity when it is determined by using a conventional method.
In particular, as introduced earlier, in the case of Newtonian fluids it is a straightforward step to determine their viscosity by performing a linear fit of $Ln\big[A(\tau)\big]$ vs. $\tau$, which here is executed for ordinate values ranging from $0$ and $-1$ (equivalent to $A(\tau)=1$ and $A(\tau)=e^{-1}$, respectively) to minimize the error.
In Figure \ref{fig:MAPE}--(A) we report the mean absolute percentage error of the fluids' viscosity evaluated as mentioned above \textit{versus} the measurement duration, which varies from $1$s to $1024$s.
The simulated trajectories were generated for optically trapped particles suspended into four fluids having viscosity spanning three orders of magnitude (i.e., from $0.001$ to $1$Pa$\cdot$s), but with all the other inputs having the following values: trap stiffness of $0.25\mu$N/m, particle radius of $1\mu$m, constant temperature of $19^o$C and acquisition rate of $1$kHz.
From the diagram it can be seen that, for short measurement duration (i.e., at $T_m=1$s) all the measurements return an error as high as circa $33\%$; whereas, as the length of the measurement increases, the MAPE decreases down to a remarkable value of only $1\%$ at $T_m=1024$s for the fluid with the lowest viscosity value of $0.001$Pa$\cdot$s; whereas, for the other fluids it would have required `significantly' longer measurements to reach a similar accuracy, as elucidated hereafter.
Notably, when the same data shown in Figure \ref{fig:MAPE}--(A) are drawn against $De^{-1}_{OT,Nom.}\propto T_m$, all the four curves collapse onto a master curve, as shown in Figure \ref{fig:MAPE}--(B).
Thus, corroborating the concept introduced earlier that ``\textit{the higher the fluid's viscosity, the longer the measurement must be}''.
This is further confirmed by the results shown in Figure \ref{fig:MAPE}--(C), which reports the outcomes of the same analysis as the one described above, but performed for a range of trap strengths varying from $0.01$ to $5\mu$N/m. From Figure \ref{fig:MAPE}--(C) it can be seen that, at relatively low trap strengths, the MAPE of the viscosity decreases as the $T_m$ increases, and all the MAPE data collapse onto the same master curve shown in Figure \ref{fig:MAPE}--(B). However, at relatively high trap strengths, the error increases again, becoming almost independent by the duration of the measurement.
Notably, this phenomenon can be explained in terms of the relative value assumed by the time-dependent fluid's shear compliance ($J(t)$) to that of the OT ($J_{OT}$) within the experimental time-window.
In particular, for a Newtonian fluid, the shear compliance assumes the following simple expression \cite{Xu1998,Tassieri2012}:
\begin{equation}
    J(\tau)=\frac{\tau}{\eta} = \frac{\langle \Delta r^2(\tau)\rangle_{t_0}\pi a}{k_BT}
\end{equation}
where $\langle \Delta r^2(\tau)\rangle_{t_0}$ is the MSD as introduced in Equation \ref{eqn:NMSD} and represented in Figure \ref{fig:Positional Data}--(D) by the sections of the MSD curves having slope $\sim1$ at relatively short lag-times, for the same combinations of fluids' viscosity and trap stiffness discussed above.
Whereas, the compliance of the OT is time-independent and it is inversely proportional to the trap strength: $J_{OT}=6\pi a/\kappa\propto \langle r^2\rangle_{eq.}$ (whose value is also represented in Figure \ref{fig:Positional Data}--(D) by the plateau values of the MSD).
It follows that, the characteristic time of the compound system $\tau_{OT}$ is identified by the lag-time at which the above two compliances are equal to each other.
Therefore, given that the accuracy to which the viscosity is calculated depends on the number of data points of the NPAF (or equivalently of MSD) available at lag-times $\tau < \tau_{OT}$ -- i.e., within the time--window ranging from $\tau_1=1/f$ (for which $A(\tau_1)\sim 1$) to $\tau=\tau_{OT}$ (for which $A(\tau_{OT})=e^{-1}$), used for the fitting procedure -- the analysis of the particles' trajectory will return viscosity values with a high degree of uncertainty at relatively large $\kappa$ values, for which $\tau_{OT}\xrightarrow[]{}\tau_1$.
Indeed, as shown in Figure \ref{fig:Positional Data} (D), for each fluid's viscosity, the effective time--window $[t_1,\tau_{OT}]$ shortens as the trap stiffness increases.
From a physics prospective, this is simply because the stronger is $\kappa$, the smaller is the particle variance from the trap centre (i.e., Eq. \ref{eqn:EquiEn}), thus overshadowing the fluid's contribution to the particle dynamics.

\begin{figure*}[t!]
    \centering
    \includegraphics[trim={0 0 0 0},clip,width=\textwidth]{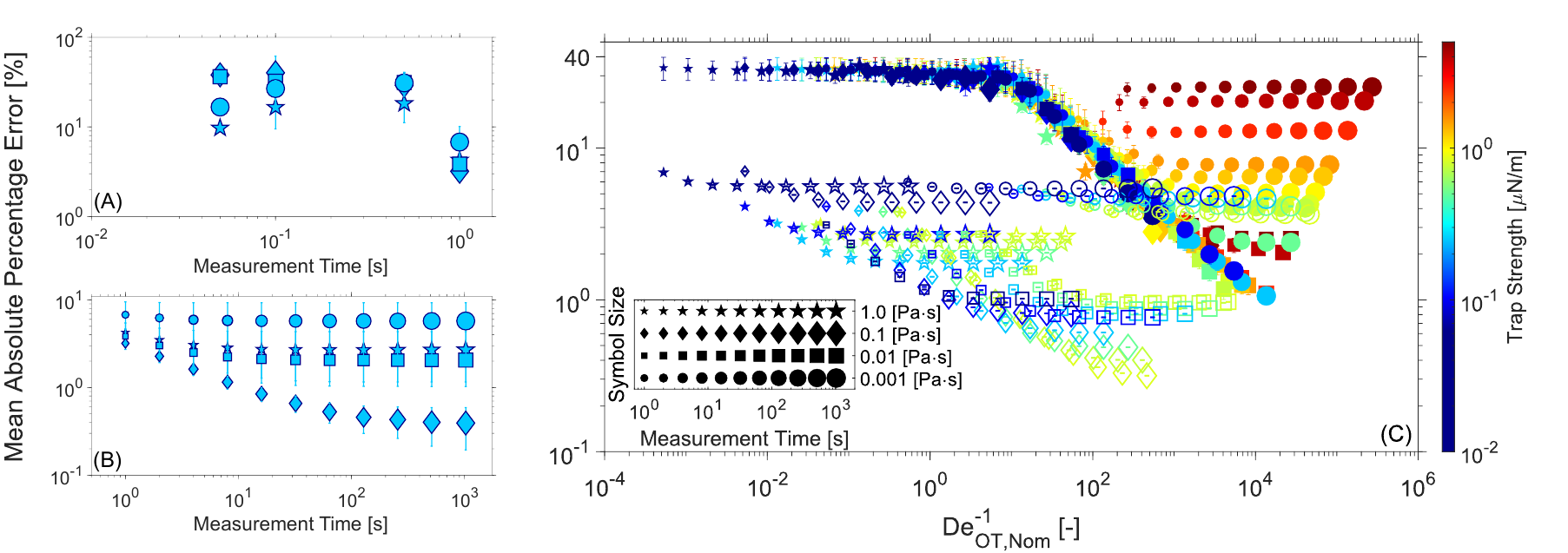}
    \caption{(A) The mean absolute percentage error (MAPE) of viscosity prediction \textit{vs}. measurement duration ($T_m$) determined by averaging the prediction error from ML algorithms with different input dimensions. (B) The MAPE of viscosity prediction averaged for 3 models with 1s input dimension \textit{vs}. measurement time. (C) The MAPE \textit{vs}. $De^{-1}_{OT,Nom.}$ for the conventional analysis shown in \ref{fig:MAPE} with an overlay for the optimal ML algorithm with an input measurement time of $1$s (open symbols). All three diagrams have the same ordinate. The colour bar indicates the nominal trap stiffness used during the generation of the trajectories. The trap stiffness range used to generate the ML curves has a narrower range due to computational constraints.}
    \label{fgr:MAPEML}
\end{figure*}

In order to better understand the \textit{optimal modus operandi} of MOT measurements, it is important to analyse the relative position of the system's characteristic time within the `finite' experimental time--window. This concept has been recently introduced by Tassieri \textit{et al.}\cite{Tassieri2018} while testing the efficacy of a novel analytical tool (i-Rheo~\textit{GT}) for converting the time--dependent materials' shear relaxation modulus into their frequency--dependent complex shear modulus. In particular, they introduced a dimensionless parameter $T_a = log(\tau/t_1)/log(t_N/t_1)$ that accounts for the relative position of the material’s characteristic relaxation time $\tau$ to that of the experimental time window $[t_1,t_N]$; where $t_1$ is the shortest time of the experimental data set (here $t_1=1/f$) and $t_N$ is the longest one (here $t_N\equiv T_m=N/f$). 
Interestingly, in the context of this work, $T_a$ assumes the following form:
\begin{equation}
    T_a = \frac{log(f\tau_{OT})}{log(N)},
    \label{Equ:TaNumber}
\end{equation}
and by plotting the MAPE of the viscosity \textit{vs}. $T_a$, as shown in Figure \ref{fig:TaNumber}, it is possible to identify a value of $T_a\simeq 1/3$ where MRAE assumes a minima. Notably, this could be used to express $N$ as a function of $De_{OT}$ via Equations \ref{eqn:De} and \ref{Equ:TaNumber}, i.e.: $N\simeq De_{OT}^{-3/2}$; thus, providing a means of estimating the number of data points to be acquired to achieve a MAPE of $\sim1\%$ for any generic fluid.
This is indeed possible if the trap stiffness of the OT rig is calibrated first in water and it is also assumed not to vary significantly when measurements are performed on different fluids (i.e., when the refractive index of the sample under investigation does not differ significantly from that of water). With these conditions satisfied, one could write:
\begin{equation}
    N\simeq N_w\eta_r^3,
    \label{Equ:N}
\end{equation}
where $N_w$ is the number of positional data points acquired during the microrheology measurement performed in water (e.g., here $N_w\simeq 10^6$) and $\eta_r=\eta/\eta_w$ is the relative viscosity of the fluid under investigation to that of water, of which an estimate is needed to determine $N$.

Moreover, from Figure \ref{fig:TaNumber}, it is interesting to notice that (i) for $T_a<1/3$, the MAPE of the viscosity decreases with a power law of $T_a^{-6}$ as $T_a\xrightarrow{}1/3$ from the left, and that (ii) the data points adhering to this scaling law are mostly those obtained from trajectories drawn by using as inputs a low viscosity value (i.e., $0.001$Pa$\cdot$s) and relatively high trap stiffness as the outcomes diverge from the minima. Thus implying that the MAPE of the viscosity in this region is governed mainly by the trap strength.
Whereas, for $T_a>1/3$, the MAPE of the viscosity follows a power law of $T_a^{3}$ and the data points adhering to this scaling law are mainly related to those trajectories drawn with a relatively low trap stiffness and relatively high viscosity, for which the measurement duration is not long enough for the bead to explore the whole potential well. 

In summary, we can argue that microrheology with OT requires long measurement times with many individual readings to achieve fluid's viscosity measurements with an error of only a few percents, which in practice translates to a measurement duration of the order of tens of minutes when dealing with fluids having viscosity close to that of water, OT rigs working at kHz and exerting a trap stiffness of the order of a few $\mu$N/m. Notably, when attempting microrheology measurements of fluids with significantly higher viscosity than water and with the same experimental conditions mentioned above, Equation \ref{Equ:N} reveals that the measurement duration would become soon `\textit{unachievable}' because $T_m$ scales with the cubic power of the relative viscosity: $T_m\simeq T_{m,s}\eta_r^3$. These conclusions further corroborate Tassieri's `\textit{opinion}' \cite{Tassieri2015_MOT} that conventional passive microrheology measurements with OT of living systems ``\textit{are not an option}'', as biological processes occur at much shorter time scales than the required $T_m$ and therefore their rheological properties could not be considered `\textit{time invariant}' during the measurements.
Thus the aim of this paper to employ machine learning algorithms to significantly shorten the duration of microrheology measurements performed with OT, as elucidated hereafter.

\subsection*{Enhanced MOT with Machine Learning}

Let us now investigate the efficacy of ML algorithms when used to enhance the accuracy of viscosity measurement of Newtonian fluids in passive MOT measurements. It would be prudent here to highlight the change in language that will occur when discussing the aforementioned ML algorithms. Indeed, throughout the previous sections, the attainment of Newtonian viscosity by means of conventional analytical methods presented in Equations \ref{eqn:Langevin}~-~\ref{eqn:NMSD} has been justifiably described as `\textit{calculated}', however the ML algorithms described in this paper specifically `\textit{predict}' the viscosity of the Newtonian fluid in question and therefore they will be described as such here.

As for the results described in Figure \ref{fig:MAPE}, the simulated trajectories used for evaluating the ML models were generated for optically trapped particles suspended into four fluids having viscosity spanning three orders of magnitude (i.e. from $0.001$ to $1$Pa$\cdot$s).
Figure \ref{fgr:MAPEML}-(A) shows the MAPE of the fluids' viscosity prediction versus the measurement time, associated with input segment length, for the ML algorithms fed with the following inputs: the trap strength, particle radius, temperature and acquisition rate having values of $0.25\mu$N/m, $1\mu$m, $19^o$C and $1$kHz, respectively. It can be seen that for measurement times shorter than $1$s, using the architecture described previously, the MAPE is as high as $40\%$ depending on fluid viscosity.
Interestingly, for a measurement time of $0.05$s the MAPE for the highest viscosity analysed of $\eta=1$Pa$\cdot$s is $10\%$, which is four times lower than the conventional method using $1$s of trajectory. This is a striking result, considering that the characteristic time for that particular point, $\tau_{OT} \approx 75$, is around 1500 times larger than the measurement time.   
Whereas, for $T_m=1$s, the prediction error drops to between $3-6\%$ across the three decades of fluid viscosity explored. Notice that, the input measurement times used in this study did not exceed a value of $1$s because of the demanding computational processes involved in training of ML algorithms. Therefore, in order to obtain consistent predictions of fluid viscosity, to extrapolate to $1024$s, the input measurement time used in Figure \ref{fgr:MAPEML}-(B) was $1$s. The extrapolation was carried out by feeding $1$s segments of particle trajectory into each of the three $1$s input ML models trained and averaging each of the predictions over increasingly longer times. The diagram shows the MAPE of the ML viscosity prediction versus the measurement time extrapolated to $1024$s using the same parameters described in Figure \ref{fgr:MAPEML}-(A). Generally, as the measurement time increases, the MAPE, starting at values between $3-6\%$, quickly drops to a plateau value for each viscosity reaching as low as $0.4\%$ for a viscosity of $0.1$Pa$\cdot$s. When compared to the conventional method in Figure \ref{fig:MAPE}-(A), the viscosity prediction errors displayed in Figure \ref{fgr:MAPEML}-(B) are significantly lower for most of the time window explored, apart from the MAPE of the conventional approach at the longest times.
It is important to highlight that, in machine learning algorithms the individual model accuracy is determined by the model hyper-parameters as well as the size and quality of training data. Moreover, the random initialisation of the training process can cause the model to learn to predict particular viscosity ranges more accurately than others. The variability in performance of different instances of the same model for different viscosity values is indicated by the error bars in Figures \ref{fgr:MAPEML}-(A-B). Notably, the significant reduction in MAPE from the conventional approach to the ML prediction occurs over the entire range of explored viscosity.

As for standard ML studies, we have selected the best performing $1$s model to be analysed for a range of trap strengths as shown in Figure \ref{fgr:MAPEML}-(C). Here, the MAPE of both the conventional method and the ML model (open symbols) are plotted versus $De^{-1}_{OT,Nom.}$ for trap strengths ranging from $0.01-5\mu$N/m. Notice that, the range of trap strengths used in ML analysis is $0.01-0.85\mu$N/m, which is slightly wider than the range of trap strengths used in training ($0.08-0.39\mu$N/m).
Figure \ref{fgr:MAPEML}-(C) shows that the MAPE values of the ML algorithm are $5$ times smaller than those of the conventional method for $De^{-1}_{OT,Nom.}<1$; i.e., $\sim7\%$ and $\sim35\%$ respectively.
It is also apparent that, unlike the conventional method, the ML error curves do not collapse into a master curve when drawn against $De^{-1}_{OT,Nom.}$.
This is believed to be due to the design of the feature extraction component of the ML architecture, which uses convolutional filters that learn local temporal structures common to both short and long trajectories. Therefore, once the model has learned to extract low- and high-dimensional local features in the measurements \textit{a priori} during the training process, the CNN can decode the fluids' viscosity `\textit{directly}' from the raw measurements using a statistically relevant number of steps $N'$ required to disambiguate features that are present in the data, rather than from statistically averaged quantities over $N=T_mf$ steps used in the standard approach. Notably, the number $N'$ can be much smaller than $N$ and no longer needs to satisfy the scaling governed by the Deborah number on the individual measurement -- as the missing information has been encoded before the measurement into the learned CNN parameters.

Therefore, ML has the ability to enhance the accuracy of passive MOT measurements by significantly reducing the measurement time from tens of minutes down to $1$s with a prediction error that is $5$ times smaller than the conventional analytical method applied to the same data. Additionally, the ML algorithm shown here is able to predict the viscosity of a Newtonian fluid across $3$ decades range and we expect that a less generalised model, which is trained on a smaller span of viscosity values, could further improve the performance of the ML approach.

\section*{Conclusion}
In this article we provide an experimental evidence supporting the observation made by Tassieri \cite{Tassieri2015_MOT} in 2015, that conventional linear microrheology with optical tweezers may not be an appropriate experimental methodology for studying the viscoelastic properties of living systems.
In particular we have focused on the analysis of computer simulated trajectories of an optically trapped particle suspended within a set of Newtonian fluids having viscosity values spanning three decades, i.e. from $10^{-3}$ to $1$ Pa$\cdot$s.
The conventional statistical mechanics analysis of these simulations has led to the following key findings: (i) we corroborate the requirement for MOT studies to perform ``\textit{sufficiently}'' long measurements when using conventional analytical methods for data analysis and (ii) we provide, for the first time in literature, a means for estimating the required duration of the experiment to achieve an uncertainty as low as $1\%$; (iii) we provide evidence explaining why conventional MOT measurements commonly underestimate the materials' viscoelastic properties, especially in the case of high viscous fluids or soft-solids such as gels and cells.
Moreover, we have developed a machine learning algorithm that uses feature extraction on only `one second' of trajectory data to determine the viscosity of Newtonian fluids, yet capable of returning viscosity values carrying an error as low as $\sim0.3\%$ at best, which is five times smaller than those obtained from conventional analytical methods applied to the same data.
Our results clearly indicate that machine learning is a valid option to be exploited to perform fast and accurate microrheology measurements with optical tweezers in living systems.

\section*{Author Contributions}

\textbf{Matthew G. Smith}: conceptualization, investigation, methodology, formal analysis, writing – original draft preparation, writing - review and editing; \textbf{Jack Radford}: investigation, visualization, formal analysis, writing - review and editing; \textbf{Helen O'Mahony}: conceptualization, investigation; \textbf{Jorge Ramírez, Eky Febrianto, Graham M. Gibson, Andrew B. Matheson}: writing - review and editing; \textbf{Daniele Faccio}: supervision, writing - review and editing; \textbf{Manlio Tassieri}: conceptualization, methodology, validation, supervision, project administration, writing – review and editing.

\section*{Conflicts of interest}
There are no conflicts to declare.

\section*{Acknowledgements}
This work was supported by the EPSRC CDT in ``Intelligent Sensing and Measurement'' (EP/L016753/1). M.T. acknowledges support via EPSRC grant ``Experiencing the micro-world - a cell's perspective'' (EP/R035067/1 – EP/R035563/1 – EP/R035156/1).

%%%END OF MAIN TEXT%%%


\balance


%%%REFERENCES%%%
\bibliography{References_ML} %You need to replace "rsc" on this line with the name of your .bib file
\bibliographystyle{ieeetr} %the RSC's .bst file

\end{document}